\documentclass[aps,prb,showpacs,twocolumn]{revtex4-1}
\usepackage{graphics,wrapfig,times}
\usepackage{graphicx}
\usepackage{amsmath}
\usepackage{amssymb}

\newcommand{\bvec}[1]{{\mathbf #1}}

\newcommand{\be}{\begin{equation}}
\newcommand{\ee}{\end{equation}}
\newcommand{\bea}{\begin{eqnarray}}
\newcommand{\eea}{\end{eqnarray}}

\usepackage{color}

\begin{document}

\title{Instabilities of a birefringent semi-metal}
\author{Nazanin Komeilizadeh and Malcolm P. Kennett}  
\affiliation{Physics Department, Simon Fraser University, 8888 University Drive, Burnaby, British Columbia, V5A 1S6, Canada}
\date{\today}

\begin{abstract}
Birefringent fermions arise as massless fermionic low energy excitations of a particular tight binding model for spinless
fermions on a square lattice which have two ``speeds of light'' [M. P. Kennett, {\it et al.}, Phys. Rev. A {\bf 83}, 053636 (2011)].  
We use mean field theory to study phases that can arise when there are nearest neighbour and next-nearest neighbour repulsive interactions in this model
and demonstrate robustness of the birefringent semi-metal phase in the presence of weak interactions and identify transitions to staggered density and quantum
anomalous Hall ordered phases. We consider the effect of coupling birefringent fermions to a magnetic field, and find
analytic expressions for the corresponding Landau levels and demonstrate that their integer Quantum Hall effect displays additional plateaux 
beyond those observed for regular Dirac fermions, such as in graphene.  We briefly discuss a tight-binding construction that leads to three dimensional
birefringent fermions.
\end{abstract}

\pacs{71.10.Fd, 37.10.Jk, 05.30.Fk, 71.10.Pm}

\maketitle
\section{Introduction}

There has been recent intense experimental and theoretical activity focused on systems with low energy excitations
with Dirac dispersions, such as Graphene, \cite{graphene} topological insulators, \cite{TI} and Weyl semi-metals.\cite{Weylsm}
Birefringent fermions are massless fermions which differ from Dirac fermions in that they
have more than one distinct velocity.  It has recently been shown that they can arise as 
the low energy excitations of a specific tight binding model \cite{Kennett} and are one 
of a class of recently investigated birefringent Dirac systems, in which 
there may be multiple Fermi velocities and/or flat bands.\cite{Bercioux,Shen,Apaja,Green,Kennett,Goldman,Lan1,Moessner,Igor,Lan2,Watanabe,Roy,Vigh}
The most promising venue for realizing such physics appears to be using cold atoms in optical 
lattices.\cite{Bercioux,Shen,Apaja,Kennett,Goldman,Lan1,Lan2}

The recent demonstration of artificial Dirac systems, in cold atoms,\cite{Tarruell}
``molecular graphene''\cite{Gomes} and dielectric resonators\cite{Bellec} opens the door to engineering Dirac-like bandstructures
and exploring their properties.  This motivates our study of birefringent fermions as an example of a system 
that generalizes regular Dirac fermions.  These fermions break the chiral $SU(2)$ symmetry present
for Dirac fermions,\cite{Hint} but do so without generating a mass,\cite{Kennett} unlike the usual case for Dirac fermions.\cite{Miransky}  
The price that is paid is that the emergent low energy Lorentz symmetry is also broken, and so one has a 
situation where there are two different Fermi velocities (or ``speeds of light''). 
In detail, the low energy theory of birefringent fermions consists of four component massless
fermions with two separate Fermi velocities $v_0 (1 \pm \beta)$ controlled by
the parameter $0 \leq \beta \leq 1$.  Writing the low energy theory in
Dirac form, the parameter $\beta$ multiplies terms in the kinetic energy not present
in the regular Dirac Hamiltonian.
 We have considered the 
response of these fermions to a variety of perturbations, and in the presence of 
topological defects\cite{Kennett,Roy}  and found that the property of birefringence is quite robust.  
An important question to ask is whether this birefringence is robust in the presence of interactions, and
the nature of broken symmetry phases that gap the birefringent semi-metal for strong enough interactions.

We consider this question here by treating repulsive interactions at a mean-field level in the previously 
introduced tight-binding model of spinless fermions 
which has birefringent fermions as its low energy excitations. \cite{Kennett}
The tight binding model we consider in fact interpolates between a
model of regular Dirac fermions on a square lattice and the Lieb lattice, which has attracted
considerable attention itself recently.\cite{Franz1,Tsai,Iglovikov}  We find that generically
the birefringent semi-metal phase is stable to weak interactions.  For sufficiently 
strong nearest neighbour interactions there is an instability to a staggered density phase 
and this tendency is enhanced as birefringence increases in strength.  For sufficiently strong 
next-nearest neighbour interactions, there can be a topologically insulating 
quantum anomalous Hall phase, which is robust to weak nearest neighbour interactions.

This paper is structured as follows.  In Sec.~\ref{sec:biref} we
recall the model of birefringent Dirac fermions. In Sec.~\ref{sec:interactions} 
we study the phases that arise due to both nearest neighbour and next-nearest
neighbour interactions.  In Sec.~\ref{sec:landau} and Sec.~\ref{sec:threed} we
consider the effects of magnetic field and the generalization of birefringent fermions
to three dimensions respectively.  Finally, in Sec.~\ref{sec:conc} we conclude and discuss
our results.

\section{Birefringent fermions}
\label{sec:biref}
\begin{figure}[htb]
\includegraphics[width=6cm]{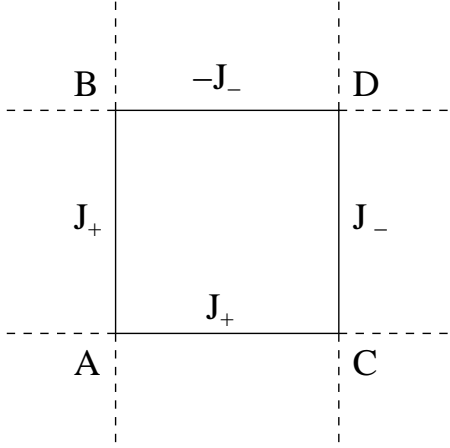}
\caption{Unit cell of tight binding model}
\label{fig:tight}
\end{figure}

We recently introduced birefringent fermions as the low energy excitations of 
the tight binding model of non-interacting spinless
fermions on a square lattice at half-filling illustrated in Fig.~\ref{fig:tight}.\cite{Kennett}  
This tight-binding model can also be viewed as 
corresponding to a model with positive hopping parameters and half a flux quantum through each 
plaquette, similar to a square lattice model considered by Seradjeh {\it et al.}\cite{Seradjeh} which admits
Dirac fermions as low energy excitations.

The dispersion relation reads as 
\begin{equation} 
 E_{k} = \pm 2J_\pm \sqrt{\cos^2 k_x + \cos^2 k_y} , 
\label{eq:biref}
\end{equation}
(the factor of 2 in Eq.~(\ref{eq:biref}) corrects Ref.~\onlinecite{Kennett})
where $J_\pm = J_0 (1 \pm \beta)$, with $0 \leq \beta \leq 1$.  This
leads to four equivalent Dirac points at the corners of the
Brillouin zone: $\bvec{K}_{\pm,\pm} = 
\left(\pm\frac{\pi}{2},\pm\frac{\pi}{2}\right)$.
Labelling the four sites in the unit cell as $A$, $B$, $C$, and $D$ we can write the low
energy theory in the form:
\begin{eqnarray}
\label{model}
H = \sum_{\bvec{k}} \psi_k^\dagger[E_k - H_k] \psi_k ,
\end{eqnarray}
where $ \psi_k^T = (c_{Ak}, c_{Bk}, c_{Ck}, c_{Dk}),$
with $c_{Ik}$ a fermionic annihilation operator for
a fermion with momentum $k$ which resides on sites $I = A, B, C,$ or $D$,
and (setting $2J_0 = 1$) 
\begin{eqnarray}
H_k & = & \left[i\left(\gamma_0 \gamma_1 + i\beta\gamma_3\right)k_x
+ i\left(\gamma_0 \gamma_2 + i\beta \gamma_5\right)k_y\right]
\label{eq:birefHam} \\
& = & H_0 + H_\beta , \nonumber
\end{eqnarray}
where $H_0$ is the Hamiltonian when $\beta = 0$ and $H_\beta$ contains all terms
involving $\beta$.
We use a non-standard representation of the gamma matrices in
which $\gamma_0 = \sigma_3 \otimes \sigma_3$, $\gamma_1 = \sigma_2
\otimes I_2$, $\gamma_2 = \sigma_3 \otimes \sigma_2$,
$\gamma_3 = -\sigma_1\otimes I_2$,
and $\gamma_5 = \gamma_0 \gamma_1 \gamma_2 \gamma_3  = -\sigma_3\otimes \sigma_1$.
The matrices $\gamma_0$, $\gamma_1$, $\gamma_2$, $\gamma_3$ and $\gamma_5$
satisfy the Clifford algebra $\gamma_\mu \gamma_\nu + \gamma_\nu
\gamma_\mu = 2\delta_{\mu\nu}$.\cite{footnote} 
The representation is four dimensional, the minimal dimension for a time-reversal invariant system of spinless
Dirac fermions in two dimensions on a lattice.\cite{Herbut} 
There are four Dirac points, however unlike graphene, 
for which the minimal representation is constructed with two sublattice degrees of freedom
and two inequivalent Dirac points, in the birefringent model the four Dirac points are equivalent and arise from there
being four lattice points in the unit cell. 
The spectrum $$E_k = \pm (1\pm \beta)|\bvec{k}|,$$ can be obtained very simply by noting that $H_0$ and $H_\beta$
commute and that both $H_0$ and $H_\beta$ individually represent Dirac Hamiltonians (albeit
for different representations of the gamma matrices). The two Dirac cones with Fermi velocities $1 \pm \beta$
make it tempting to think that it might be possible to use a 
direct sum with two copies of two component massless
Dirac fermions or Weyl fermions.  This is not possible without breaking time reversal 
symmetry,\cite{nielsen,haldane} which is respected here.\cite{Herbut}
In the limit $\beta = 1$ the model is analagous 
to the previously studied case of the Lieb 
lattice\cite{Shen,Apaja,dagotto} and there are two flat bands at zero 
energy and a two component Weyl fermion.

\section{Interactions}
\label{sec:interactions}
We focus on nearest neighbour and 
next-nearest neighbour repulsive interactions introduced via the Hamiltonian

\begin{equation} 
          H_{\rm int}= \sum_{i,j} V_{ij} \hat{n}_i \hat{n}_j ,
\end{equation} 
where we note that for spinless fermions there can be no on-site interactions.  
We study the low energy theory in the vicinity of one of the Dirac cones at the corners of the Brillouin zone and ignore
scatterings between different Dirac cones.
We represent the generating functional as an imaginary time 
path integral over Grassmann-valued fields $\psi$ and $\bar{\psi}= \psi^\dagger \gamma_0$: 
\begin{equation} 
 \mathcal{Z} = \displaystyle\int [ \mathcal{D}\bar{\psi}{\mathcal D}\psi ]  e^{-S[\bar{\psi},\psi]} ,
\end{equation} 
where $S[\bar{\psi},\psi] =  \displaystyle\int^{\beta}_{0} d\tau \, {\mathcal  L}(\bar{\psi},\psi)$ is the action
and the Lagrangian ${\mathcal L}$ is 
\begin{eqnarray}
{\mathcal L} & = & {\mathcal L}_0 + {\mathcal L}_\beta + {\mathcal L}_{\rm int}  ,
\end{eqnarray}
with ${\mathcal L}_0$ the Lagrangian associated with $H_0$, ${\mathcal L}_\beta$ the Lagrangian associated with $H_\beta$ 
and ${\mathcal L}_{\rm int}$ the interaction Lagrangian.
  We treat the interactions at a mean field level, similarly to approaches previously used for graphene\cite{Hint,HJR}
by solving the saddle point equations for the order parameters obtained from the path integral formalism.

\subsection{Nearest Neighbour interactions}
We first consider nearest neighbour interactions with strength $V_1$.  
We can decouple the quartic interaction terms in the action by introducing Hubbard-Stratonovich fields and making
use of the identities corresponding to Hartree and Fock decompositions of
$$ n_A n_B + n_A n_C +n_B n_D + n_C n_D,$$
which are written out explicitly in Appendix~\ref{sec:ident1}. 

In principle we should introduce Hubbard-Stratonovich fields corresponding to all Hartree and Fock decompositions of the interaction
term, but we find that for nearest neighbour interactions the leading instability is to staggered density order with order 
parameter
$$ \left<\chi\right> \propto - \left<\bar{\psi}\psi\right> = \left<n_B\right> + \left<n_C\right> - \left<n_A\right> - \left<n_D\right>.$$
Keeping only the Hartree term [Eq.~(\ref{eq:nnhartree})] and also introducing the field $$ \phi \propto n_A + n_B + n_C + n_D,$$
we get
\begin{eqnarray} 
 \mathcal{Z} &=& \int [\mathcal{D} \chi][\mathcal{D} \phi] e^{-S[\chi]-S[\phi]} \int [\mathcal{D} \bar{\psi}][{\mathcal D}\psi] 
e^{-{S_0}[\bar{\psi},\psi]-S_{\rm mix}} , \nonumber \\ & & 
\end{eqnarray} 
where 
\begin{eqnarray}
S[\chi] & = & \frac{1}{2V_1}  \int^\beta_0 d\tau  \int d^2\vec{x} \, [\chi(\vec{x}, \tau)]^2, \nonumber \\
S[\phi] & = & \frac{1}{2V_1}  \int^\beta_0 d\tau  \int d^2\vec{x} \,  [\phi(\vec{x}, \tau)]^2, \nonumber \\
S_0[\bar{\psi}, \psi]  & = & \int^\beta_0 d\tau \left[  {\mathcal L}_{0} + {\mathcal L}_\beta [\bar{\psi},\psi] \right], \nonumber
\end{eqnarray}
and 
\begin{eqnarray*} 
S_{\rm mix}& =&   \displaystyle\int^\beta_0 d\tau  \displaystyle\int d^2\vec{x} \, [\chi \bar{\psi}\psi + i \phi \bar{\psi}\gamma_0\psi ] .
\end{eqnarray*} 
After integrating out the Grassman fields, we may write the generating functional as
\begin{eqnarray*} 
\mathcal{Z} 
&=& \displaystyle\int [\mathcal{D} \chi][\mathcal{D}\phi] \displaystyle e^{-S[\chi]-S[\phi] +{\rm Tr}(\ln M)} , 
\end{eqnarray*} 
where after Fourier transforming,
\begin{eqnarray*} 
M &=& i \gamma_\mu k_\mu -  \beta\gamma_0\gamma_3 k_1  -\beta \gamma_0\gamma_5 k_2 + \chi+ i\phi\gamma_0 .
\end{eqnarray*} 
We take a saddle point approximation:
\begin{eqnarray*} 
\frac{\left<\chi\right>}{V_1}  =  {\rm Tr}[M^{-1}], \quad\quad
\frac{\left<\phi\right>}{V_1} = {\rm Tr}(i \gamma_0 M^{-1}) \nonumber ,
\end{eqnarray*} 
and find that the saddle point equation for $\chi$ (where $\Lambda$ is an ultra-violet cutoff) gives the critical interaction
strength, $V_c$ as

\begin{figure}[htb]
\includegraphics[width=4.5cm,angle=270]{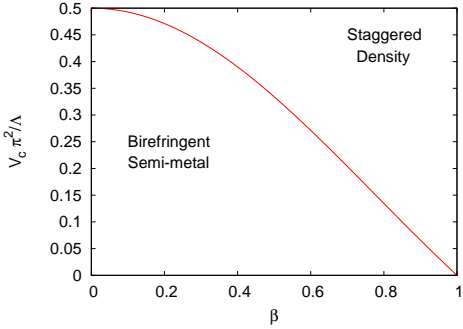}
\caption{Phase diagram in the presence of nearest neighbour interactions 
as a function of $V_1$ and $\beta$.}
\label{fig:chibeta}
\end{figure}

\begin{widetext}
\begin{eqnarray}
\frac{1}{V_c} & = & \frac{1}{2\pi^2} \int_{-1}^1 dx \int_0^\Lambda dk \left\{ \frac{k^2}{\chi^2 + k^2[(1-\beta)^2 + x^2\beta(2-\beta)]}
 + \frac{k^2}{\chi^2 + k^2[(1+\beta)^2 - x^2 \beta (2+\beta)]}\right\}.
\end{eqnarray}
We find that $\phi =0$ and that a non-zero solution for $\chi$ may be found provided $V_1 \geq V_c$, where (keeping only terms that 
scale with $\Lambda$) we can obtain $V_c$ as a function of $\beta$:
 \begin{eqnarray} 
\frac{1}{V_c} &=& 
\frac{\Lambda}{ \pi^2} \left\{ \frac{1}{(1 - \beta) \sqrt{\beta(2-\beta)}}\tan^{-1}\left(\frac{\sqrt{\beta(2-\beta)}}{1-\beta}\right) 
+ \frac{1}{2(1+\beta)\sqrt{\beta(2+\beta)}}\ln \left|\frac{1+ \frac{1+\beta}{\sqrt{\beta(2+\beta)}}}{1-\frac{1+\beta}{\sqrt{\beta(2+\beta)}}}\right|\right\} ,
 \end{eqnarray} 
which is illustrated in Fig.~\ref{fig:chibeta}.  The result reduces to the previously calculated expression in the 
limit $\beta \to 0$.\cite{Kaveh}

\end{widetext}

Figure 2 illustrates that as $\beta$ approaches unity, the birefringent semi-metal becomes increasingly more susceptible to 
staggered density order.   In the presence of a non-zero staggered density order parameter, $\chi$, the spectrum of the 
birefringent fermion model becomes gapped $$\epsilon_{\pm\pm} = \pm \sqrt{\chi^2 + 4J_\pm^2 k^2},$$ but retains its
birefringent property.\cite{Kennett} The spectrum 
reduces to that found with a usual mass term when $\beta = 0$.

\subsection{Next Nearest neighbour interactions}
\begin{figure}[ht]
\includegraphics[width=6cm]{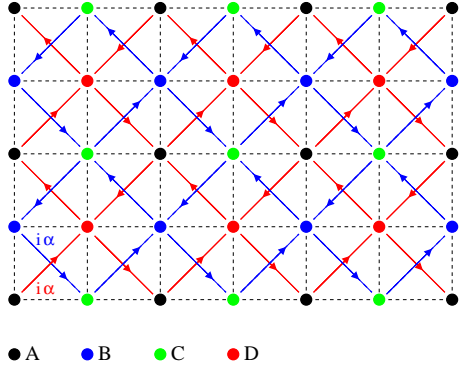}
\caption{Hopping pattern corresponding to non-zero $\zeta_{35}$ order parameter.  The amplitude of 
all diagonal hopping integrals is identical, and equal to $\alpha$.}
\label{fig:circ}
\end{figure}

\begin{figure}
\includegraphics[width=7cm]{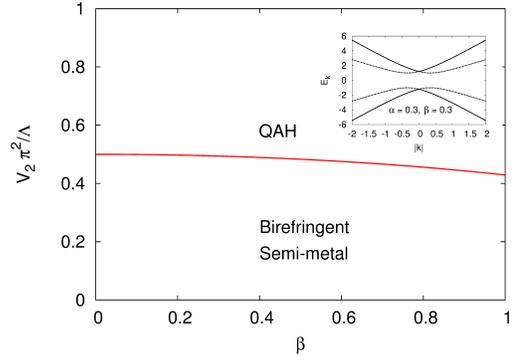}
\caption{Phase diagram as a function of $\beta$ and $V_2$ for next nearest neighbour interactions, showing the
birefringent semi-metal phase and the Quantum anomalous Hall (QAH) phase.  The inset shows the dispersion for
$\alpha = 0.3$ and $\beta = 0.3$.}
\label{fig:nnn}
\end{figure}

In order to deal with next nearest neighbour interactions
(with strength $V_2$) we again focus in the vicinity of a single Dirac cone, and make use of decompositions of 
$n_A n_D + n_B n_C$ in Hartree and Fock channels (full details in Appendix~\ref{sec:ident2}).
We hence introduce additional Hubbard Stratonovich fields to decompose this interaction term in the 
path integral.  We find that for $V_1 = 0$, the leading instability is to the order parameter 
\begin{eqnarray}
\zeta_{35}  & \propto & \left<\bar{\psi}i\gamma_3\gamma_5\psi\right> \nonumber  \\
 & = & -i\left[\left<\psi^\dagger_A \psi_D\right> - \left<\psi^\dagger_D \psi_A\right> +
 \left<\psi^\dagger_B \psi_C\right> - \left<\psi^\dagger_C \psi_B\right>\right], 
\nonumber
\end{eqnarray}
which breaks time reversal symmetry \cite{haldane} and leads to 
an additional term in the action:

\begin{eqnarray}
S_{\rm mix}^{\rm nnn} & = & \int_0^\beta d\tau \int d^2\bvec{x}  \, \zeta_{35} \bar{\psi}i\gamma_3\gamma_5 \psi ,
\end{eqnarray}
which has a critical coupling
\begin{widetext}
\begin{eqnarray}
\frac{1}{V_{35}} = \frac{2\Lambda}{\pi^2} \frac{1}{\beta^2 (\beta^2 - 4)} \frac{1}{\gamma_+ + \gamma_-} \left\{
\frac{(1 - \beta^2 + \beta^2 \gamma_+)}{2\sqrt{\gamma_+}}\ln\left|\frac{1 - \sqrt{\gamma_+}}{1 + \sqrt{\gamma_+}}\right|
 - \frac{(1 - \beta^2 -\beta^2 \gamma_-)}{\sqrt{\gamma_-}} \tan^{-1}\left(\frac{1}{\sqrt{\gamma_-}}\right)\right\},
\end{eqnarray}
where
 \begin{eqnarray} 
\gamma_\pm = \pm \frac{ (\beta^2-3)}{ (\beta^2-4)} + 
\displaystyle\sqrt{\displaystyle\frac{(\beta^2-3)^2}{ (\beta^2-4)^2}  + \displaystyle\frac{ (1-\beta^2)^2}{ \beta^2(4-\beta^2)}} .
\end{eqnarray}
\end{widetext}
At a mean field level, this ordering is equivalent to having a circulating current in the pattern 
illustrated in Fig.~\ref{fig:circ}, giving a quantum anomalous Hall phase similar to that found
on the honeycomb lattice\cite{Raghu} or in the three band Hubbard model for cuprates\cite{Weber} when next nearest
neighbour interactions are present.
If the order is as illustrated in Fig.~\ref{fig:circ}, with an amplitude $\alpha$, then the dispersion 
is always gapped and rotationally invariant and takes the form 

\begin{eqnarray}
E_k & = & \pm \sqrt{2(J_+^2 + J_-^2)|\bvec{k}|^2 + 16\alpha^2 \pm  2D_k} \\
  D_k &= & \sqrt{(J_+^2 - J_-^2)^2|\bvec{k}|^4 + 16(J_+^2 - J_-^2)\alpha^2 |\bvec{k}|^2} \nonumber .
\end{eqnarray}
For non-zero $\alpha$, there is always a gap of $8\alpha$ between the upper and lower bands, but 
for non-zero $\beta$, the minimum gap is for a ring of finite $k$, with radius
$$k = \frac{2\sqrt{2} \alpha}{\sqrt{J_+^2 - J_-^2}}\left(\sqrt{1 + \frac{J_+^2 - J_-^2}{2J_-^2}} - 1\right).$$
The dispersion is shown as an insert in Fig.~\ref{fig:nnn}, which also illustrates the phase diagram 
when only next nearest neighbour interactions are present.

Now, as noted in Ref.~\onlinecite{Kennett}, a different representation of the gamma matrices can transform
$\gamma_0\gamma_1 \leftrightarrow \gamma_3$ and $\gamma_0\gamma_2 \leftrightarrow \gamma_5$, essentially
swapping $H_0$ and $H_\beta$ (up to a factor of $\beta$) in Eq.~(\ref{eq:birefHam}).  The same transformation
takes $\zeta_{35}$ to 
\begin{eqnarray} 
 \zeta_{12}  & \propto & \left<\bar{\psi}i\gamma_1\gamma_2\psi\right> \nonumber  \\
 & = & i\left[\left<\psi^\dagger_A \psi_D\right> - \left<\psi^\dagger_D \psi_A\right> -
 \left<\psi^\dagger_B \psi_C\right> + \left<\psi^\dagger_C \psi_B\right>\right], \nonumber 
\end{eqnarray}
which also breaks time reversal symmetry and
in which the direction of the circulating current on either the AD or BC sublattice is reversed 
with respect to $\zeta_{35}$ ordering.  If $\beta > 1$ then $\zeta_{12}$ ordering is favoured and 
for $1< \beta < 2$, the critical coupling is

\begin{widetext}
\begin{eqnarray}
\frac{1}{V_{12}} 
 = \frac{2\Lambda}{\pi^2} \frac{1}{\beta^2 (\beta^2 - 4)} \frac{1}{\gamma_+ + \gamma_-} \left\{
\frac{(1 - \beta^2 +(2-\beta^2) \gamma_-)}{\sqrt{\gamma_-}} \tan^{-1}\left(\frac{1}{\sqrt{\gamma_-}}\right)
 -\frac{(1 - \beta^2 - (2-\beta^2) \gamma_+)}{2\sqrt{\gamma_+}}\ln\left|\frac{1 - \sqrt{\gamma_+}}{1 + \sqrt{\gamma_+}}\right| \right\}, 
\nonumber \\
\end{eqnarray}
\end{widetext}
which tends to the same value as  $V_{35}$ in the limit $\beta \to 1$, at which $H_0$ and $H_\beta$ have
equal weight in the Hamiltonian.

\subsection{Nearest neighbour and next nearest neighbour interactions}
When both $V_1 \neq 0$ and $V_2 \neq 0$, there can be either $\chi$ or $\zeta_{35}$ ordering.  
We note that these two orders are even and odd respectively under the discrete symmetry operator
$\Gamma$ that was introduced in Ref.~\onlinecite{Kennett}, which in Euclidean form is
$$ \Gamma = \frac{i}{2}\left(\gamma_2\gamma_3 + \gamma_1\gamma_5\right) - \frac{i}{2}\left(\gamma_1\gamma_3 
+ \gamma_2\gamma_5\right),$$
and on the lattice corresponds to a reflection about the diagonal $AD$ in the
unit cell, with $c_A \to c_A$, $c_B \to c_C$, $c_C \to c_B$ and
$c_D \to -c_D$. The action of $\Gamma$ on the Hamiltonian, $H_k$,  is to exchange $k_x$ and $k_y$.
The effects on $\gamma_0$, $\gamma_3\gamma_5$, and $\gamma_1\gamma_2$ are
$$ \Gamma \gamma_0 \Gamma = \gamma_0, \quad \quad \Gamma \gamma_3 \gamma_5 \Gamma = - \gamma_3\gamma_5, \quad \quad \Gamma \gamma_1 \gamma_2 \Gamma = - \gamma_1\gamma_2 .$$
We calculate the phase diagram for $V_1 \neq 0$ and $V_2 \neq 0$  below.

\begin{figure}[h]
\includegraphics[width=4.5cm,angle=270]{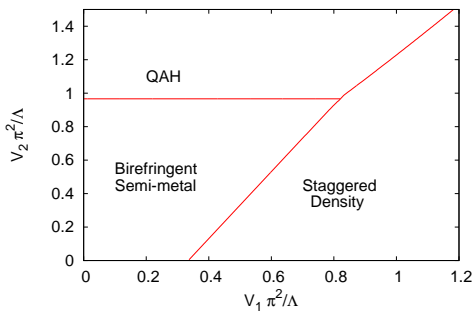}
\caption{Phase diagram as a function of $V_1$ and $V_2$ at 
$\beta =0.5$}
\label{fig:v1v2}
\end{figure}

Note that the coupling strength at which the transition from semi-metal to QAH phase occurs differs 
between Figs.~\ref{fig:nnn} and \ref{fig:v1v2} because of a factor of 2 in the action.  
In calculating the phase diagram shown in Fig.~\ref{fig:nnn} we assume that there are no other forms of
ordering when performing the decomposition into Hubbard-Stratonovich fields, and in Fig.~\ref{fig:v1v2} we
decompose the next-nearest neighbour interaction terms equally between $\chi$ and $\zeta_{35}$ ordering.  The qualitative behaviour
of the phase diagram shown in Fig.~\ref{fig:v1v2} is maintained for all $\beta$ but the exact positions 
of the phase transition lines have some $\beta$ dependence which can be anticipated 
from Figs.~\ref{fig:chibeta} and \ref{fig:nnn}.

\section{Landau Levels}
\label{sec:landau}

\begin{figure}[ht]
\begin{center}
\includegraphics[width=4.5cm,angle=270]{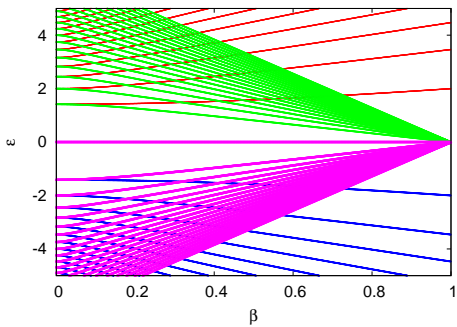}
\end{center}
\caption{Landau level energy eigenvalues as a function of $\beta$ for $n=0$ to $n=20$.}
\label{fig:landaulevels}
\end{figure}

Whilst we originally derived birefringent fermions in a tight binding model with an artificial magnetic field, \cite{Kennett}
we are free to ask what the spectrum of birefringent fermions looks like when a magnetic field is present, without asking about 
specific tight binding models that might be required to realize them.  Hence, we now derive the Landau level energy spectrum
by coupling a magnetic field to the effective Hamiltonian.  We write the Hamiltonian in real space, and use minimal coupling
and the Landau gauge $\bvec{A} = (0, Bx,0)$.  Noting that the system is translationally invariant in the $y$-direction we
make the following ansatz for the eigenstates: $\psi^T (x,y) = e^{ik_y y} (f_1(x), f_2(x), f_3(x), f_4(x)),$
and after the change of variables $ \xi = (k_y + B x)/\sqrt{B}$, $\epsilon = E/2J_0\sqrt{B}$ we obtain the eigenvalue
equation
\begin{widetext}
 \begin{equation}
 \left( \begin{array}{cccc}
\epsilon & -(1+\beta)\xi & i (1+\beta)\displaystyle\frac{\partial}{\partial \xi} & 0 \\
(1+\beta)\xi & -\epsilon & 0 & i(1-\beta)\displaystyle\frac{\partial}{\partial \xi}  \\
-i(1+\beta)\displaystyle\frac{\partial}{\partial \xi}  & 0 &  -\epsilon & (1-\beta)\xi \\
0 & i(\beta-1)\displaystyle\frac{\partial}{\partial \xi} & (\beta-1)\xi & \epsilon  \end{array} \right) \left( \begin{array}{cccc}
f_1(\xi) \\
f_2(\xi)  \\
f_3(\xi) \\
f_4(\xi)
  \end{array} \right) =0 .
\end{equation}
We can combine the equations above to obtain the following differential equation for $f_1$:
\begin{eqnarray}
 \left[\epsilon^2 - (1-\beta)^2 \xi^2 + (1-\beta)^2 \frac{\partial^2}{\partial \xi^2}\right] 
\left[\epsilon^2 - (1+\beta)^2 \xi^2 + (1+\beta)^2 \frac{\partial^2}{\partial \xi^2}\right] f_1(\xi) - (1-\beta^2)^2 f_1(\xi) = 0 .
\end{eqnarray}

Using the ansatz $f_1(\xi) = H_n(\xi) \exp\left[ -\xi^2/2\right]$, with $H_n(\xi)$ the $n^{\rm th}$ Hermite polynomial,
we arrive at the following expression for the energy eigenvalues
$$ \epsilon_{n,\beta,\pm\pm} = \pm \sqrt{(2n+1)(1+\beta^2) \pm \sqrt{(2n+1)^2 (1+\beta^2)^2 - 4n (n+1) (1-\beta^2)^2}}.$$
To connect to the standard relativistic Landau levels, first consider $\beta = 0$, which gives 
$ \epsilon_n = \pm \sqrt{2n+2},$ or $ \epsilon_n = \pm \sqrt{2n}.$ When $\beta =0$, the system retains the $SU(2)$ chiral symmetry
generated by $\{\gamma_3,\gamma_5,\gamma_3\gamma_5\}$ and the Landau levels are doubly degenerate.  When $\beta$ is non-zero,
the chiral symmetry is broken, and the degeneracy is lifted so that as $\beta \to 1$, half of the 
levels go to $\epsilon = 0$ and the other half to 
 $\epsilon = \pm \sqrt{2(2n+1)}$ as illustrated in Fig.~\ref{fig:landaulevels}.  

\end{widetext}

\subsection{Integer Quantum Hall effect}
In graphene the integer Quantum Hall effect shows plateaux at $\sigma = \pm (4n+2)e^2/h$,
which is a result of the four fold degeneracy (two from spin and two from valley degrees
of freedom) for $n\neq 0$ Landau levels and two fold degeneracy of the $n=0$ Landau level.\cite{Novoselov,Sharapov}
For spinless fermions as considered here, when $\beta = 0$ one would expect to see an 
integer quantum Hall effect with $\sigma = fe^2/h$ with $f = \pm(2n+1)$, since there is 
no factor of 2 associated with spin degeneracy.  When $\beta \neq 0$ the breaking of the 
degeneracy of the Dirac cones implies that the integer Quantum Hall effects will also be 
modified so that conductivities for all non zero integers should be present, i.e.
$\sigma = fe^2/h$ with $f = \pm (n+1)$. 

\section{Birefringent fermions in three dimensions}
\label{sec:threed}
\begin{figure}[ht]
\includegraphics[width=7cm]{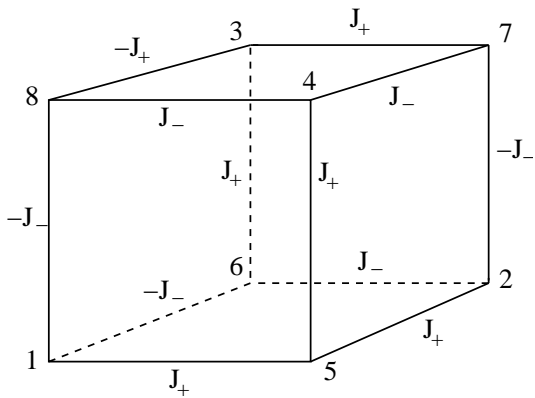}
\caption{Eight site unit cell and hopping parameters 
for the three dimensional birefringent fermion model.}
\label{fig:3dbiref}
\end{figure}

All of our discussions of birefringent fermions have focused on two dimensions, but it is interesting
to ask whether this physics is realisable in three dimensions as well.  Hosur {\it et al.} considered
a staggered flux model in three dimensions, which has Dirac fermions as its low energy excitations. 
By choosing the same flux pattern and allowing for both $J_+$ and $J_-$ hopping amplitudes in the 
8 site unit cell as illustrated in Fig.~\ref{fig:3dbiref}, one can obtain birefringent fermions.  

Introduce an eight-component fermion operator at momentum $\vec{k}$: 
$f^\dagger_{i\bvec{k}} = (A^\dagger_{1\bvec{k}}, A^\dagger_{2\bvec{k}}, A^\dagger_{3 \bvec{k}},A^\dagger_{4 \bvec{k}},
A^\dagger_{5\bvec{k}}, A^\dagger_{6\bvec{k}},A^\dagger_{7 \bvec{k}},A^\dagger_{8\bvec{k}})$, and then the  
tight binding Hamiltonian can be represented as
$$H = \displaystyle\sum_k f^\dagger_{i\bvec{k}} H_{ij\bvec{k}} f_{j\bvec{k}},$$ 
with the non-zero elements of the hopping matrix, $H_k$ equal to: \\
$H_{15}= 2J_+ \cos k_x$, $H_{16} = -2J_- \cos k_y$, $H_{18} = -2J_- \cos k_z$, 
$H_{25}= 2J_+ \cos k_y$, $H_{26} = 2J_- \cos k_x$, $H_{27} = -2J_- \cos k_z$, 
$H_{36}= 2J_+ \cos k_z$, $H_{37} = 2J_+ \cos k_x$, $H_{38} = -2J_+ \cos k_y$, 
$H_{45}= 2J_+ \cos k_z$, $H_{47} = 2J_- \cos k_y$, $H_{48} = 2J_- \cos k_x$. 
This may be written  in terms of the Pauli matrices in the following form
\begin{eqnarray*} 
H_k &=& 2J_0\left[ \cos k_x (\sigma_1 \otimes I_2 \otimes I_2) + \beta \cos k_x (\sigma_1 \otimes I_2 \otimes \sigma_3)\right.\nonumber\\
&+&  \cos k_y (\sigma_2 \otimes I_2 \otimes \sigma_2) +  \beta \cos k_y (\sigma_1 \otimes \sigma_3 \otimes \sigma_1)\nonumber\\ 
& +& \left. \cos k_z (\sigma_2 \otimes \sigma_2 \otimes \sigma_1) +  \beta \cos k_z (\sigma_1 \otimes \sigma_1 \otimes \sigma_1) \right]\nonumber\\ 
 \end{eqnarray*} 
Hence, the energy eigenvalues are given by
$$E_k = \pm 2J_\pm\displaystyle\sqrt{\cos^2 k_x + \cos^2 k_y + \cos^2 k_z},$$
and at the eight vertices of the Brillouin zone $\bvec{K}_{\pm,\pm,\pm} = \left(\pm \frac{\pi}{2}, \pm \frac{\pi}{2}, \pm \frac{\pi}{2}\right)$,
there are Dirac points at which the dispersion takes the form $E_\pm,\pm = \pm 2J_\pm |k|$, and each of the birefringent fermion
bands is doubly degenerate.

\section{Discussion and Conclusions}
\label{sec:conc}
In conclusion, we have studied the effects of interactions within the tight binding model 
for birefringent fermions at a mean-field level.  We have illustrated that for nearest neighbour and next-nearest
neighbour repulsive interactions on the square lattice,  the birefringent 
semi-metal persists up to a critical interaction strength for $\beta < 1$.  We have calculated the
$\beta$ dependence of these critical interaction strengths and find that as $\beta$ approaches
1, the system is more susceptible to interactions than for small values of $\beta$.  A staggered density
phase is favoured by nearest neighbour interactions and a quantum anomalous Hall phase with circulating
currents is favoured by next-nearest neighbour interactions.  
The ordered phases that arise for birefringent
fermions are thus quite similar to those of regular Dirac fermions, but as $\beta$ is increased,
birefringent fermions have a lower critical interaction strength for ordering.  It should be noted that 
in the case of the honeycomb lattice, despite mean field predictions of quantum anomalous Hall phases,\cite{Franz1,Raghu,Grushin}
such phases have proven to be less robust in exact diagonalization calculations with periodic boundary
conditions \cite{Dag,Garcia}, but present when open boundary conditions are used.\cite{Duric}  We expect
that similar considerations apply to the situation considered here and consider the exploration of 
interaction effects beyond mean field theory to be an interesting avenue for future work.

Despite similarities at zero magnetic field, birefringent fermions display qualitatively different behaviour to regular Dirac fermions in 
the presence of a magnetic field.  We considered the effects of a magnetic
field on the spectrum of birefringent fermions and obtained exact expressions for their Landau levels.  
The broken chiral symmetry of birefringent fermions lifts the degeneracy of Landau levels for 
regular Dirac fermions, and hence there are additional integer Quantum Hall plateaux compared to 
the case when $\beta = 0$.

The study presented here is part of the broader effort of understanding interaction effects in 
novel bandstructures.  Future directions to consider include
the addition of spin to birefringent fermions, which would also allow for on-site Hubbard interactions.
The combination of interactions and magnetic field would also be interesting to investigate.  The 
experimental realization of birefringent fermions would also be of great interest -- we proposed a
scheme to realize them in a cold atom setting in Ref.~\onlinecite{Kennett} for cold atoms, but given the
nature of the tight binding models for two and three dimensional birefringent fermions, it is conceivable
that they might arise naturally in transition metal compounds where $d$-orbitals are important for 
hopping matrix elements.

\section{Acknowledgements}
We acknowledge helpful discussions with  Kamran Kaveh, Chi-Ken Lu and Bitan Roy
and in particular wish to thank Igor Herbut for encouragement and 
insightful suggestions. We also thank Matthew Fitzpatrick for a close reading of the 
manuscript.  This work was supported by NSERC.

\begin{appendix}
\begin{widetext}
\section{Useful Identities}
\label{sec:ident}
For convenience, we list here the identities we used in decomposing interaction terms into 
Hartree and Fock channels.

\subsection{Nearest Neighbour interactions}
\label{sec:ident1}
We can decompose nearest neighbour interaction terms into Hartree and Fock channels using the 
following identities: for the Hartree decomposition:
\begin{eqnarray} 
n_A n_B + n_A n_C +n_B n_D + n_C n_D &=& \displaystyle\frac{1}{4}[ (n_A + n_B +n_C +  n_D)^2 - (n_A - n_B - n_C +  n_D)^2 ] \nonumber\\ 
&=& \displaystyle\frac{1}{4}\left[ \left(\bar{\psi}\gamma_0\psi\right)^2 - \left( \bar{\psi}\psi\right)^2\right]
\label{eq:nnhartree} 
\end{eqnarray} 
and the Fock decomposition:
\begin{eqnarray}
n_A n_B + n_A n_C + n_B n_D + n_C n_D & = & -\frac{1}{8} \left\{ \left(\bar{\psi}\gamma_0 \gamma_1 \psi\right)^2 
 + \left(\bar{\psi}i\gamma_1 \psi\right)^2 + \left(\bar{\psi}\gamma_0 \gamma_3\psi\right)^2 + 
\left(\bar{\psi}i\gamma_3\psi\right)^2 \right. \nonumber \\ & & \left. \hspace*{1cm}
 + \left(\bar{\psi}\gamma_0\gamma_2\psi\right)^2  + \left(\bar{\psi} i\gamma_2\psi\right)^2 
+ \left(\bar{\psi}\gamma_0\gamma_5\psi\right)^2 + \left(\bar{\psi}i\gamma_5\psi\right)^2\right\},
\end{eqnarray}

\subsection{Next-Nearest Neighbour Interactions}
\label{sec:ident2}
The decompositions for next-nearest neighbour interactions are
\begin{eqnarray}
n_A n_D + n_B n_C & = & \frac{1}{8} \left[ (n_A + n_B + n_C + n_D)^2 + (n_A - n_B - n_C + n_D)^2 \right. \nonumber \\ & & \left.
 - (n_A - n_B + n_C - n_D)^2 - (n_A + n_B - n_C - n_D)^2\right] \nonumber \\
& = & \frac{1}{8} \left[ \left(\bar{\psi} \gamma_0 \psi\right)^2 + \left(\bar{\psi}\psi\right)^2 - \left(\bar{\psi} i\gamma_1\gamma_3\psi\right)^2
 + \left(\bar{\psi}i\gamma_2\gamma_5\psi\right)^2\right] ,
\end{eqnarray}
for the Hartree channel and
\begin{eqnarray}
n_A n_D + n_B n_C & = & - \frac{1}{8} \left[ \left(\bar{\psi}i\gamma_1\gamma_5\psi\right)^2 + \left(\bar{\psi}i\gamma_2\gamma_3\psi\right)^2
+\left(\bar{\psi}i\gamma_3\gamma_5\psi\right)^2 + \left(\bar{\psi}i\gamma_1\gamma_2\psi\right)^2\right]
\end{eqnarray}
for the Fock channel.
\end{widetext}

\end{appendix}


\begin{thebibliography}{99}

\bibitem{graphene} K.S. Novoselov, A. K. Geim, S. V. Morozov, D. Jiang, Y. Zhang, S. V. Dubonos,
I. V. Grigorieva, and A. A. Firsov, Science {\bf 306}, 666 (2004); G. W. Semenoff,
 Phys. Rev. Lett. {\bf 53}, 2449 (1984).

\bibitem{TI} C. L. Kane and E. J. Mele, Phys. Rev. Lett. {\bf 95}, 146802 (2005);
D. Hsieh, D. Qian, L. Wray, Y. Xia, Y. S. Hor, R. J. Cava, and M. Z. Hasan, Nature {\bf 452}, 970 (2008).

\bibitem{Weylsm} S. M. Young, S. Zaheer, J. C. Y. Teo, C. L. Kane, E. J. Mele, and A. M. Rappe,
  Phys. Rev. Lett. {\bf 108}, 140405 (2012).

\bibitem{Kennett} M. P. Kennett, N. Komeilizadeh, K. Kaveh, and P. M. Smith, Phys. Rev. A {\bf 83}, 053636 (2011).
\bibitem{Bercioux} D. Bercioux, D. F. Urban, H. Grabert, and W. H\"{a}usler, Phys. Rev. A {\bf 80}, 063603 (2009).
\bibitem{Shen} R. Shen, L. B. Shao, B. Wang, and D. Y. Xing, Phys. Rev. B {\bf 81}, 041410(R) (2010).
\bibitem{Apaja} V. Apaja, M. Hyrk\"{a}s, and M. Manninen, Phys. Rev. A {\bf 82}, 041402(R) (2010).
\bibitem{Green} D. Green, L. Santos, and C. Chamon, Phys. Rev. B {\bf 82}, 075104 (2010).
\bibitem{Goldman} N. Goldman, D. F. Urban, and D. Bercioux, Phys. Rev. A {\bf 83}, 063601 (2011).
\bibitem{Lan1} Z. Lan, N. Goldman, A. Bermudez, W. Lu, P. Ohberg,  Phys. Rev. B {\bf 84}, 165115 (2011).
\bibitem{Moessner} B. D\'{o}ra, J. Kailasvuori, and R. Moessner, Phys. Rev. B {\bf 84}, 195422 (2011).
\bibitem{Igor}  Chi-Ken Lu and Igor F Herbut, J. Phys. A: Math. Theor.{\bf 44},  295003 (2011).
\bibitem{Lan2} Z. Lan, A. Celi, W. Lu, P. Ohberg, and M. Lewenstein, Phys. Rev. Lett. {\bf 107}, 253001 (2011).
\bibitem{Watanabe} H. Watanabe, Y. Hatsugai, and H. Aoki, J. Phys.: Conf. Ser. {\bf 334}, 012044 (2011).
\bibitem{Roy} B. Roy, P. M. Smith, and M. P. Kennett, Phys. Rev. B {\bf 85}, 235119 (2012). 
\bibitem{Vigh} M. Vigh, L. Oroszl\'{a}ny, S. Vajna, P. San-Jose, G. D\'{a}vid, J. Cserti, and B. D\'{o}ra, Phys. Rev. B
 {\bf 88}, 161413(R) (2013).

\bibitem{Tarruell} L. Tarruell, D. Greif, T. Uehlinger, G. Jotzu, T. Esslinger,  Nature {\bf 483}, 302 (2012).
\bibitem{Gomes} K. K. Gomes, W. Mar, W. Ko, F. Guinea, and H. C. Manoharan, Nature {\bf 483}, 306 (2012).
\bibitem{Bellec} M. Bellec, U. Kuhl, G. Montambaux, and F. Mortessagne, Phys. Rev. B {\bf 88}, 115437 (2013).

\bibitem{Hint} I. F. Herbut, Phys. Rev. Lett. {\bf 97}, 146401 (2006).
\bibitem{Miransky} V. A. Miransky, {\it Dynamical Symmetry Breaking in
Quantum Field Theories} (World Scientific, Singapore, 1993).


\bibitem{Franz1} C. Weeks and M. Franz, Phys. Rev. B {\bf 85}, 041104(R) (2012).
\bibitem{Tsai} W.-F. Tsai, C. Feng, H. Yao, and J. Hu, arXiv:1112.5789v1.
\bibitem{Iglovikov} V. I. Iglovikov, F. H\'{e}bert, B. Gr\'{e}maud, G. G. Batrouni, and  R. T. Scalettar, arXiv:1404.5482v4.

\bibitem{Seradjeh} B. Seradjeh, C. Weeks, and M. Franz, Phys. Rev. B {\bf 77}, 033104 (2008).


\bibitem{footnote} In Ref.~\onlinecite{Kennett} a Minkowski rather than Euclidean
metric was used. This implies slightly modified expressions
for  $\gamma_{1,2,3,5}$.

\bibitem{Herbut} I. F. Herbut, Phys. Rev. B {\bf 83}, 245445 (2011).
\bibitem{nielsen} H. B. Nielsen and M. Ninomiya, Nucl. Phys. {\bf 185}, 20 (1981).
\bibitem{haldane} F. D. M. Haldane, Phys. Rev. Lett. {\bf 61}, 2015 (1988).
\bibitem{dagotto} E. Dagotto, E. Fradkin, and A. Moreo, Phys. Lett. {\bf 172}, 383 (1986).

\bibitem{HJR} I. F. Herbut, V. Juri\v ci\' c, and B. Roy, Phys. Rev. B {\bf 79}, 085116 (2009).

\bibitem{Kaveh} K. Kaveh and I. F. Herbut, Phys. Rev. B {\bf 71}, 184519 (2005).

\bibitem{Raghu} S. Raghu, X.-L. Qi, C. Honerkamp, and S.-C. Zhang, Phys. Rev. Lett. {\bf 100}, 156401 (2008).
\bibitem{Weber} C. Weber, A. L\"{a}uchli, F. Mila, and T. Giamarchi, Phys. Rev. Lett. {\bf 102}, 017005 (2009).

\bibitem{Novoselov} K. S. Novoselov, A. K. Geim, S. V. Morozov, D. Jiang, M. I. Katsnelson, I. V. Grigorieva, 
  S. V. Dubonos, and   A. A. Firsov, Nature {\bf 438}, 197 (2005).
\bibitem{Sharapov} V. P. Gusynin and S. G. Sharapov, Phys. Rev. Lett. {\bf 95}, 146801 (2005).

\bibitem{Hosur} P. Hosur, S. Ryu, and A. Vishwanath, Phys. Rev. B {\bf 81}, 045120 (2010).

\bibitem{Grushin} A. G. Grushin, E. V. Castro, A. Cortijo, F. de Juan, M. A. H. Vozmediano, and B. Valenzuela,
 Phys. Rev. B {\bf 87}, 085136 (2013).

\bibitem{Dag} M. Daghofer and M. Hohenadler, Phys. Rev. B {\bf 89}, 035103 (2014).

\bibitem{Garcia} N. A. Garc\'{i}a-Martinez, A. G. Grushin, T. Neupert, B. Valenzuela, and E. V. Castro, Phys. Rev. 
 B {\bf 88}, 245123 (2013).

\bibitem{Duric} T. Duri\'{c}, N. Chancellor, and I. F. Herbut, Phys. Rev. B {\bf 89}, 165123 (2014).


\end{thebibliography}
\end{document}